\documentclass[10pt,twocolumn,cleanfoot,cleanhead]{asme2e}

\usepackage{empheq, amssymb}
\usepackage{lmodern}
\usepackage{amsmath}
\usepackage{amsfonts}
\usepackage{amssymb}
\usepackage{bm} 
\usepackage{txfonts}
\usepackage[T1]{fontenc}
\usepackage[utf8]{inputenc} 
\usepackage[dvipsnames]{xcolor}
\usepackage[american]{babel} 
\usepackage{graphicx}
\usepackage{epstopdf}
\usepackage[noadjust]{cite} 
\usepackage{ltxcmds}
\usepackage{mathtools}
\usepackage{subcaption}
\usepackage{multirow} 
\usepackage{blindtext}
\PassOptionsToPackage{hyphens}{url}
\usepackage{hyperref}
\usepackage{enumitem}
\usepackage{ifthen}
\usepackage{etoolbox}
\usepackage{cancel}
\usepackage{empheq}
\usepackage{fontawesome}
\usepackage{soul}
\usepackage{flushend} 
\usepackage{textcomp}
\usepackage{lipsum}

\definecolor{light-gray}{gray}{0.4}
\definecolor{box-gray}{gray}{1}

\hypersetup{
    unicode=false,          % non-Latin characters in Acrobat’s bookmarks
    pdftoolbar=true,        % show Acrobat’s toolbar?
    pdfmenubar=true,        % show Acrobat’s menu?
    pdffitwindow=false,     % window fit to page when opened
    pdfstartview={FitV},    % fits the width of the page to the window
    pdftitle={Integrated Design for Wave Energy Converter Farms: Assessing Plant, Control, Layout, and Site Selection Coupling in the Presence of Irregular Waves},    % title
    pdfauthor={Saeed Azad; Suraj Khanal; Daniel R. Herber; Gaofeng Jia},     % author
    pdfsubject={},   % subject of the document
    pdfkeywords = {}, % list of keywords
    pdfnewwindow=true,      % links in new window
    colorlinks=true,
    allcolors=blue
}

\usepackage{nomencl}
\renewcommand\nomgroup[1]{%
  \item[\bfseries
  \ifstrequal{#1}{V}{ Variables}{%
  \ifstrequal{#1}{B}{ Subscripts}{%
  \ifstrequal{#1}{P}{ Notation}{%
  \ifstrequal{#1}{A}{ Acronyms}{}}}}]
}

\makenomenclature



\usepackage{dblfloatfix}
\usepackage{float}
\definecolor{block-gray}{gray}{0.95}

\usepackage[framemethod=TikZ]{mdframed}
\usepackage{xpatch}

\makeatletter
\xpatchcmd{\endmdframed}
  {\aftergroup\endmdf@trivlist\color@endgroup}
  {\endmdf@trivlist\color@endgroup\@doendpe}
  {}{}
\makeatother

\usepackage{fontawesome}
\usepackage{titlesec}

\usepackage{hyperref}

\newcommand{\doi}[1]{{doi:~\href{http://doi.org/#1}{#1}}\rmFullStop}

\newcommand*{\rmFullStop}{\rmifnextchar{.}{}{}}

\makeatletter
\newcommand{\rmifnextchar}[3]{%
  \begingroup
  \ltx@LocToksA{\endgroup#2}%
  \ltx@LocToksB{\endgroup#3}%
  \ltx@ifnextchar{#1}{%
    \def\next{\the\ltx@LocToksA}%
    \afterassignment\next
    \let\scratch= %
  }{%
    \the\ltx@LocToksB
  }%
}
\makeatother

\usepackage{colortbl}

\definecolor{light-gray}{gray}{0.6}

\newcommand{\xsection}[1]{\section[#1]{\MakeUppercase{#1}}}
\usepackage{accents}

\definecolor{needcolor}{HTML}{C62828}

\usepackage[most]{tcolorbox} %  colored and framed text boxes
\definecolor{block-gray}{gray}{0.95}

\AtBeginDocument{%
 \abovedisplayskip=4pt plus 1pt minus 1pt
 \abovedisplayshortskip=0pt plus 3pt
 \belowdisplayskip=4pt plus 1pt minus 1pt
 \belowdisplayshortskip=0pt plus 3pt
}

\newtcolorbox{xreviewer}{%
	empty,
    borderline west = {4pt}{0pt}{gray},
    boxrule = 0pt,
    boxsep = 0pt,
    breakable,
    colback = block-gray,
    enhanced,
    frame hidden,
    left skip = 0pt,
    notitle,
    parbox = false,
    sharp corners,
}
\newtcolorbox{xresponse}{%
	empty,
    boxsep = 0pt,
    breakable,
    frame hidden,
    notitle,
    parbox = false,
}
\newtcolorbox{xchange}{%
    boxrule = 1pt,
    boxsep = 0pt,
    breakable,
    colframe = black,
    enhanced jigsaw,
    interior hidden,
    notitle,
    parbox = false,
    arc=0pt,
    outer arc=0pt,
    after skip=20pt plus 2pt,
}

\usepackage{catchfilebetweentags}
\makeatletter
\def\CatchFBT@Fin@l#1[#2]{%
   \begingroup
      \makeatletter #2%
      \scantokens\expandafter{%
         \expandafter\CatchFBT@tok\expandafter{\the\CatchFBT@tok}}%
      \CatchFBT@IsAToken{#1}
         {\global#1\expandafter{\the\CatchFBT@tok}}
         {\xdef#1{\the\CatchFBT@tok}}%
      \ifx\CatchFBT@tok#1\else\global\CatchFBT@tok{}\fi
   \endgroup
}% \CatchFBT@Final
\makeatother

\newtheorem{postulate}{Postulate}

\usepackage{multirow}
\usepackage{threeparttable}
\usepackage{booktabs}
\usepackage{tabularx}
\newcolumntype{Y}{>{\centering\arraybackslash}X}
\usepackage{tablefootnote} % for table footnotes

\newcommand{\parm}{\mathord{\color{black!33}\bullet}}%
\newcommand{\RNum}[1]{\uppercase\expandafter{\romannumeral #1\relax}}
\newcolumntype{s}{>{\columncolor[HTML]{F5F5F5}} c}
\newcommand{\MyBold}[1]{\mathbf{#1}}
\newcommand{\Rwec}{R_{\text{wec}}}            % Variable for WEC radius
\newcommand{\ARwec}{AR_{\text{wec}}}            % Variable for WEC aspect ratio
\newcommand{\Dwec}{D_{\text{wec}}}            % variable for WEC draft
\newcommand{\Rwecmax}{\bar{R}_{\text{wec}}}   % Variable for maximum WEC radius
\newcommand{\Dwecmax}{\bar{D}_{\text{wec}}}   % variable for maximum WEC draft
\newcommand{\ARwecmax}{\bar{AR}_{\text{wec}}}   % variable for maximum WEC aspect
\newcommand{\Rwecmin}{\underaccent{\bar}{R}_{\text{wec}}}   % Variable for maximum WEC radius
\newcommand{\Dwecmin}{\underaccent{\bar}{D}_{\text{wec}}}   % variable for maximum WEC draft
\newcommand{\ARwecmin}{\underaccent{\bar}{AR}_{\text{wec}}}   % variable for maximum WEC aspect
\newcommand{\Nwec}{n_{\text{wec}}}     % variable for number of WECs in the farm
\newcommand{\Mass}{\MyBold{M}}       % Mass Matrix
\newcommand{\Force}{\MyBold{F}}      % Force matrix (due to array)
\newcommand{\OMF}{(\omega)}             % omega function argument
\newcommand{\AL}{\MyBold{w}}           % Array Layout symbol
\newcommand{\KPTO}{\MyBold{K}_{\text{pto}}}           % k pto vector
\newcommand{\BPTO}{\MyBold{B}_{\text{pto}}}           % B pto vector
\newcommand{\KPTOmin}{\underaccent{\bar}{\MyBold{k}}_{\text{pto}}}           % Array Layout symbol
\newcommand{\BPTOmin}{\underaccent{\bar}{\MyBold{B}}_{\text{pto}}}           % Array Layout symbol
\newcommand{\KPTOmax}{\bar{\MyBold{k}}_{\text{pto}}}           % Array Layout symbol
\newcommand{\BPTOmax}{\bar{\MyBold{B}}_{\text{pto}}}           % Array Layout symbol
\usepackage{units}

\usepackage{tikz}
\usetikzlibrary{matrix,fit} 
\usetikzlibrary{matrix}
\usetikzlibrary{graphs}
\usepackage{pgfkeys} 
\usetikzlibrary{positioning}
\usepackage{lipsum,adjustbox}
\usetikzlibrary{arrows.meta}

\title{Integrated Design for Wave Energy Converter Farms: Assessing Plant, Control, Layout, and Site Selection Coupling in the Presence of Irregular Waves}

\author{Saeed~Azad\thanks{Corresponding author, \texttt{\href{mailto:saeed.azad@colostate.edu}{saeed.azad@colostate.edu}}} 
\affiliation{
Postdoctoral Fellow\\
Department of Systems Engineering \\
Colorado State University\\
Fort Collins, CO 80523 \\
Email:~\texttt{\href{mailto:saeed.azad@colostate.edu}{saeed.azad@colostate.edu}}\\ 
}
}

\author{Suraj~Khanal
\affiliation{
Graduate Student\\
Department of Civil and Environmental Engineering\\
Colorado State University \\
Fort Collins, CO 80523 \\
Email:~\texttt{\href{mailto:Suraj.Khanal@colostate.edu}{suraj.khanal@colostate.edu}}
}
}

\author{Daniel~R.~Herber
\affiliation{
Assistant Professor\\
Department of Systems Engineering \\
Colorado State University \\
Fort Collins, CO 80523 \\
Email:~\texttt{\href{mailto:daniel.herber@colostate.edu}{daniel.herber@colostate.edu}}
}
}

\author{Gaofeng~Jia
\affiliation{
Associate Professor\\
Department of Civil and Environmental Engineering \\
Colorado State University \\
Fort Collins, CO 80523 \\
Email:~\texttt{\href{mailto:gjia@colostate.edu}{gjia@colostate.edu}}
}
}

\begin{document}
 \setlength{\parskip}{0pt}
 \setlength{\parsep}{0pt}
 \setlength{\headsep}{0pt}
 \setlength{\topsep}{0pt}

% equations
\abovedisplayshortskip=3pt
\belowdisplayshortskip=3pt
\abovedisplayskip=3pt
\belowdisplayskip=3pt

\titlespacing*{\section}{0pt}{18pt plus 1pt minus 1pt}{3pt plus 0.5pt minus 0.5pt}

\titlespacing*{\subsection}{0pt}{9pt plus 1pt minus 0.5pt}{1pt plus 0.5pt minus 0.5pt}

\titlespacing*{\subsubsection}{0pt}{9pt plus 1pt minus 0.5pt}{1pt plus 0.5pt minus 0.5pt}

\maketitle

%---------------
\begin{abstract}\noindent
\textit{A promising direction towards reducing the levelized cost of energy for wave energy converter (WEC) farms is to improve their performance.
WEC design studies generally focus on a single design domain (e.g., geometry, control, or layout) to improve the farm's performance under simplifying assumptions, such as regular waves.
This strategy, however, has resulted in design recommendations that are impractical or limited in scope because WEC farms are complex systems that exhibit strong coupling among geometry, control, and layout domains.
In addition, the location of the candidate site, which has a large impact on the performance of the farm, is often overlooked.
Motivated by some of the limitations observed in WEC literature, this study uses an integrated design framework, based on simultaneous control co-design (CCD) principles, to discuss the impact of site selection and wave type on WEC farm design.
Interactions among plant, control, and layout are also investigated and discussed using a wide range of simulations and optimization studies.
All of the studies were conducted using frequency-domain heaving cylinder WEC devices within a farm with a linear reactive controller in the presence of irregular probabilistic waves.
The results provide high-level guidelines to help the WEC design community move toward an integrated design perspective.
}
\end{abstract}

\vspace{1ex}

%---------------
% Introduction
\xsection{Introduction}\label{sec:introduction}

Ocean waves are an important source of renewable energy that remain largely untapped.
A multitude of wave energy converter (WEC) technologies, each with different operating principles, have been developed since 1799 \cite{Ross2012}. 
The resulting technological diversity, along with challenges associated with hostile sea environment, wide variations in wave climate, and uncertainties, have thwarted the commercial development of WECs \cite{Ringwood2023}.  

\begin{table*}[t]
\caption{An overview of postulates discussed in this article.}
\label{Tab:Postulates}
\renewcommand{\arraystretch}{1.1}
\setlength{\tabcolsep}{4pt}
\centering
\begin{tabular}{c|l}
\hline  \hline
\textbf{Postulate}     & \textbf{Text} \\
\hline  
\ref{P_site} & Optimal WEC farm design depends on the selected site \\
\ref{P_Wave} & Optimal WEC farm design depends on the wave type \\
\ref{P_Plant} & Optimal WEC plant design is sensitive to variations in control, layout, and selected site \\
\ref{P_Control} & Optimal WEC control design is sensitive to variations in plant, layout,  and selected site \\
\ref{P_Layout} & Optimal WEC layout design is sensitive to variations in plant, control,  and selected site  \\
\hline \hline 
\end{tabular}
\end{table*}

With a wide range of challenges still present in the design of both WEC devices and farms, research and development efforts are the primary pathway to meaningfully reducing the cost of electricity from ocean waves.
To this end, a promising solution is to improve the technology performance level \cite{weber2012wec} of the WEC device/farm by leveraging domain coupling from the early stages of the design process.
Control co-design (CCD) \cite{GarciaSanz2019}, which refers to the simultaneous optimization of plant and control, has gained prominence in recent WEC literature \cite{herber2013wave, PenaSanchez2022, Stroefer2023, Coe2020}.
More recently, layout optimization and site selection have been considered within the CCD activity of WEC farms \cite{Azad24, Azad2023}, indicating that beyond geometry and control, the layout configuration of the farm, as well as wave variability at a given site location, are key design drivers for WEC farms.
By leveraging the coupling between plant (WEC geometry, layout, etc.) and control design, these studies have adopted a system-level mindset that has the potential to improve the performance of these complex systems.

Despite such advances, it is generally observed that the broader WEC literature often overlooks some key considerations in the design of these systems.
This approach has resulted in design recommendations that are impractical, limited in scope, or sometimes conflicting with conclusions from other WEC literature.
As an example, Lyu et al. conducted a concurrent optimization of WEC dimension and layout for arrays of $3$, $5$, and $7$ WECs using genetic algorithm (GA) \cite{Lyu2019}. 
However, only a single study with $3$ WECs was performed in irregular waves, which was truncated after $7$ GA generations due to high computational expense.
Despite the work's value in improving our understanding of WEC layout and dimensions in regular waves, the results have limited application in real-world sea states.  
Using fixed array configurations, instead of solving a layout optimization problem, also limits the performance of the farm and the extent to which conclusions from such studies can be useful \cite{Borgarino2012, Andres2014}.  
Similarly, optimization of a single domain (geometry or control or layout) \cite{Zeng2022, Moarefdoost2017, Tedeschi2012}, without accounting for the impact from the coupled domains, results in an inscrutable and/or sub-optimal design.

Motivated by the limitations in the WEC literature, this study aims to combine insights and considerations from our own investigations and the WEC literature to inform optimal decision-making in the design of WEC farms.
Anchored in the integrated design perspective of CCD, our discussions on key design considerations are supported by investigations that leverage domain interactions between WEC dimension, such as geometrical attributes (e.g.,~radius, and draft); power take-off (PTO) control (e.g.,~PTO stiffness, damping ratio, power saturation limits); and layout (device location within a farm).
Critical considerations such as wave type and site location are also discussed.
{An overview of these topics is presented in Tab.~\ref{Tab:Postulates}.} 
Through these, we aim to provide high-level guidelines to assist the WEC community in moving toward an integrated design vision that has the potential to improve the state-of-the-art WEC technology. 

The remainder of this study is organized as follows:
Sec.~\ref{sec:Background} provides some background information on model development, wave climate, assumptions, and limitations of the investigations carried out in this article.
Sec.~\ref{sec:Discussion}, which is structured in a Postulate form, discusses the impact of site location, wave type, and domain coupling among geometry, control, and layout.
Finally, conclusions are discussed in Sec.\ref{sec:conclusion}.

%---------------
%Background
\xsection{Background}\label{sec:Background}

The discussions offered in the later sections of the article are often supported by simulations or solutions from a series of optimization problems.
In this section, we briefly describe the WEC farm model, problem elements, and formulations used to develop the supporting materials.
We also discuss some of the limitations and inherent assumptions used.

With the exception of Postulate~\ref{P_Wave}, where regular waves are used for the sake of comparison, this article models the wave climate as uni-directional, irregular waves, characterized by the JONSWAP spectrum \cite{Ning2022}.
The availability of historical data for certain site locations in different regions from 1976 through 2005 facilitates the integration of site selection into our demonstrations.
Such data is used to construct the joint probability distribution of waves for various significant wave heights $H_{s}$ and wave periods $T_{p}$ \cite{Storlazzi2015, Erikson2016}.
Further details regarding wave climate and modeling can be found in our previous work \cite{Azad2023}.

We consider a WEC farm, with $\Nwec$ heaving cylinder devices, characterized by two plant optimization variables $\bm{p} \in \mathbb{R}^{2}$, namely the WEC radius $\Rwec$, and its slenderness ratio (ratio of radius to draft) $\ARwec$, such that $\bm{p} = [\Rwec, \ARwec]^{T}$.
Using linear potential flow theory, the WEC farm is modeled in the frequency domain, with the following equation of motion:
\begin{align}
\label{eqn:EquationofMotion1}
     -{\omega}^{2} \Mass \hat{\bm{\xi}}\OMF = \hat{\Force}_{\text{FK}}\OMF + \hat{\Force}_{\text{s}}\OMF + \hat{\Force}_{\text{r}}\OMF +  \hat{\Force}_{\text{hs}}\OMF + \hat{\Force}_{\text{pto}}\OMF
\end{align}

\noindent
where $\hat{\parm}$ is the complex amplitude of $\parm$, $\hat{\bm{\xi}}(\cdot) \in \mathbb{R}^{\Nwec \times 1}$ is the displacement vector, $\hat{\Force}_{\text{FK}}(\cdot)$ is the Froude-Krylov force, $\hat{\Force}_{\text{s}}(\cdot)$ is the scattering force vector, $\hat{\Force}_{\text{r}}(\cdot)$ is the radiation force, $ \hat{\Force}_{\text{hs}}(\cdot)$ is the hydrostatic force, $\hat{\Force}_{\text{pto}}(\cdot)$ is the power-take-off (PTO) force, all defined in $\in \mathbb{R}^{\Nwec \times 1}$.
In addition, $\Mass \in \mathbb{R}^{\Nwec \times \Nwec}$ is the diagonal mass matrix.
The linear PTO force is calculated as function of $\BPTO$, and $\KPTO$, which constitute the control optimization variables $\textbf{u} = [\BPTO,\KPTO]^{T} \in \mathbb{R}^{2}$.
Note that we assume uniform PTO variables across the farm. 
For any farm studies with a total of $n_\text{wec}$ devices, the layout optimization variable $\AL$ is a $2\times (n_\text{wec}-1)$ vector comprising of the coordinates of the center of each body in the Cartesian coordinate system, with the first WEC always at the origin of the coordinate system.
The candidate location of the site is described by $\bm{L}$ and affects power generation through the joint probability distribution of the waves.

The estimation of hydrodynamic coefficients within an optimization problem is often a computational bottleneck that limits the complexity of WEC problems that can be investigated.
For example, solving a concurrent layout optimization problem using genetic algorithm (GA) with multiple scattering (MS) \cite{ohkusu1974hydrodynamic, Goeteman2022} can take between $20$ to $30$ days on a high-performance computer.
Recent studies indicate that advanced methods based on surrogate modeling in combination with many-body expansion principles are promising in reducing this computational cost \cite{Azad2023, Azad24}. 
Nevertheless, in this article, we use MS to isolate our conclusions from many of the errors and uncertainties inherent in surrogate modeling by keeping the size of the optimization problems manageable and ensuring that they are computationally tractable.

Unless otherwise stated, all of the optimization problems investigated with irregular waves use an objective function $p_{v}$ that considers the average power per unit volume of the WEC device \cite{falnes2020ocean}, over $30$ years of the farm operation. 
For a general case, the WEC farm optimization problem is formulated as:
\begin{subequations}
 \label{Eqn:OPtimization}
 \begin{align}
 \underset{\bm{p},\bm{u}, \AL}{\textrm{minimize:}}
 \quad & - p_{v}(\bm{p}, \bm{u}, \AL)   \label{Eqn:Obj} \\
 \textrm{subject to:} \quad
    \begin{split}
    &  2\Rwec + s_{d} - \bm{L}_{pq} \leq 0 \quad\\
         & \qquad \forall ~~  p,q = 1, 2, \dots, \Nwec \quad p \neq q \label{Eqn:distanceconst}
    \end{split} \\
    & P_{\text{matrix}} \leq p_{\text{limit}}~\text{(optional)} \label{Eqn:PowerSaturation} \\
    & \Dwecmin \leq \Dwec \leq \Dwecmax \label{Eqn:draftcons} \\
    &  \underaccent{\bar}{\bm{p}} \leq \bm{p} \leq \bar{\bm{p}}\label{Eqn:plantconst}\\
    &  \underaccent{\bar}{\bm{u}} \leq \bm{u} \leq \bar{\bm{u}} \label{Eqn:controlconst}\\
    &  \underaccent{\bar}{\AL} \leq \AL \leq \bar{\AL}\label{Eqn:layoutconst}\\
 \textrm{where:} \quad & \bm{p} = [\Rwec, \ARwec]^{T} \in \mathbb{R}^{2} \notag\\
                & \bm{u} = [\KPTO, \BPTO]^{T} \in \mathbb{R}^{{2}}  \notag \\
                & \AL = [\bm{x},\bm{y}] \in \mathbb{R}^{{2}(\Nwec-1)} \notag 
                % & \Dwec = \Rwec/\ARwec \notag 
 \end{align} 
\end{subequations}

\noindent
where $\underaccent{\bar}{\parm}$, and $\bar{\parm}$ are lower and upper bounds, respectively, and $\Dwec$ is the draft of WEC device.
$\bm{L}_{pq}$ is the distance between $p$th and $q$th WEC devices, and $s_{d}$ is a minimum safety distance to allow maintenance ships to pass.
Note that Eq.~(\ref{Eqn:distanceconst}) is present only for optimization problems that involve layout as one of the design variables. 
Eq.~(\ref{Eqn:PowerSaturation}), when included, imposes a power saturation limit on the device power matrix and is discussed next.
The problem parameters and their values are listed in Table~\ref{Tab:Parameter_Opt}.

 \begin{table}[t]
    \caption{Problem parameters.}
    \label{Tab:Parameter_Opt}
    \renewcommand{\arraystretch}{1.1}
    \setlength{\tabcolsep}{4pt}
    \centering
    \begin{tabular}{s l s l}
    \hline  \hline
    \textrm{\textbf{Option}} & \textrm{\textbf{Value}} &  \textrm{\textbf{Option}} & \textrm{\textbf{Value}} \\
    \hline
    $\Rwecmin$ & $0.5~[\unit{m}]$ & $\Rwecmax$ & $10~[\unit{m}]$ \\
    $\ARwecmin$ & $0.2$ & $\ARwecmax$ & $10$ \\
    $\Dwecmin$ & $0.5~[\unit{m}]$ & $\Dwecmax$ & $20~[\unit{m}]$ \\
    $\KPTOmin$ & $-5\times 10^{5}~[\unit{N/m}]$ & $\KPTOmax$ & $5\times 10^{5}~[\unit{N/m}]$ \\
    $\BPTOmin$ & $0~[\unit{Ns/m}]$ & $\BPTOmax$ & $5\times 10^{5}~ [\unit{Ns/m}]$\\
    $\underaccent{\bar}{\bm{x}}$ & $0~[\unit{m}]$ & $\rho$ & $1025~[\unit{kg/m^3}]$ \\
    $\bar{\bm{x}}$ & $0.5\sqrt{2\Nwec \times 10^4}~[\unit{m}]$ & $g$ & $9.81~[\unit{m/s^2}]$ \\
    $\underaccent{\bar}{\bm{y}}$ & $-0.5\sqrt{2\Nwec \times 10^4}~[\unit{m}]$ & $s_{d}$ & $10~[\unit{m}]$\\ 
    $\bar{\bm{y}}$ & $0.5\sqrt{2\Nwec \times 10^4}~[\unit{m}]$  & $n_{yr}$   & $30~[\unit{years}]$  \\
    \hline \hline
    \end{tabular}
\end{table}

The investigations carried out in this article often implement an optimization problem that entails control variables.
The control optimization problem, however, is not well posed in the frequency domain.
This conclusion is because an optimal control solution in the frequency domain results in high and unrealistic device motions that are above practical limits. 
Addressing this issue demands more advanced approaches such as pseudo-spectral controllers \cite{Coe2020} or latching control \cite{Babarit2006}.
These, however, fall outside of the scope of the current study.

One alternative simplified approach to address this issue is to directly impose various power saturation limits on WEC power matrices of devices within the farm.
This strategy allows us to limit the power that can be captured by the WEC devices and pose an indirect constraint on device motion.
However, this approach, which is described mathematically in Eq.~(\ref{Eqn:PowerSaturation}), may result in a flat objective function for a range of PTO design parameters. 
With this limitation on the controller design in mind, we proceed to some of the assumptions used in this study. 

All of the results and demonstrations provided in this article are conducted with the following assumptions:

\begin{enumerate}[topsep=0pt,itemsep=-1ex,partopsep=1ex,parsep=1ex,label=$\bullet$]

\item Usage of unidirectional irregular wave models, while limited in comparison to multi-directional waves, is reasonable in assessing domain dependencies in WEC farms.

\item Despite certain limitations, frequency-domain WEC models are useful in providing initial performance assessments and insights, particularly on domain coupling.

\item The WEC devices in the studies will exclusively exhibit heave motion. 

\item A linear reactive PTO control with a power saturation limit is a reasonable characterization of the control system in capturing domain interactions. 

\item Maximization of power per volume of the WEC device is a reasonable substitute for an economic objective function, such as levelized cost of energy.

\item Mooring costs and electrical interconnections are ignored.
    
\end{enumerate}

With all the background, limitations, and assumptions discussed, we use the following definition of the \textit{WEC farm design}in order to clarify the scope and intent of this article.

\begin{definitionxx*}
WEC farm design refers to the determination of the physical or geometric attributes; control architecture and parameters; and layout configuration of the WEC devices within a farm.
\end{definitionxx*}

\noindent
It is worth noting that this description is by no means comprehensive and that there are a multitude of other factors and insights, some of which can already be found in the WEC literature.
Nevertheless, our goal is to emphasize some of these design considerations to highlight their role in the design of WEC farms.

%----------------
% Section 3
\xsection{Design Considerations}\label{sec:Discussion}

This section presents the core contributions of this work and is structured in the form of postulates.
While unconventional, this structure is motivated by the clarity, conciseness, and independence offered in the postulate environment.
Specifically, the impact of site location on the design and performance of WEC farms is discussed in Postulate~\ref{P_site}.
Sensitivity of WEC farm design to wave modeling is discussed in Postulate~\ref{P_Wave}.
In Postulate~\ref{P_Plant}, we discuss and demonstrate the dependency of WEC optimal plant design to control, layout, and site location.
The coupling between optimal control solution and plant, layout, and site location is demonstrated in Postulate~\ref{P_Control}.
Finally, optimized layout coupling with plant, control, and site location is discussed in Postulate~\ref{P_Layout}.

%----------------
% P_Site

\begin{postulate}\label{P_site}
Optimal WEC farm design depends on the selected site.
\end{postulate}

While intuitive, the dependency of WEC farm design, and thus its performance, on the selected site is often nuanced. 
For a given site location, the naturally-available
wave energy resource is often available in the literature (such as Refs.~\cite{Jacobson2011, OceanEnergy}).
The technically-available wave resource, also known as a recoverable resource, while also available in the literature (such as Ref.~\cite{Jacobson2011}), depends on the current state of the technology and its progress in the near future. 
Novel, system-level WEC designs have the potential to affect the recoverable resource. 
Nevertheless, these novel designs can be affected by the naturally-available wave energy resource, particularly when an economic objective drives the design decisions. 
The interplay between the naturally-available resource and an optimal WEC design drives WECs' dependency on-site location.

\begin{figure}[t]
    \captionsetup[subfigure]{justification=centering}
    \centering
    \begin{subfigure}{0.5\columnwidth}
    \centering
    \includegraphics[scale=0.49]{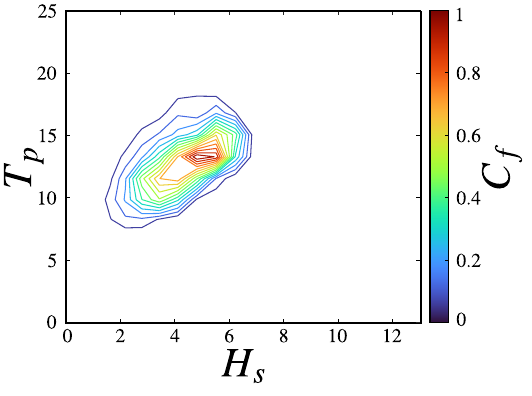}
    \caption{West Coast.}
    \label{subfig:WC_location_CF}
    \end{subfigure}%
    \begin{subfigure}{0.5\columnwidth}
    \centering
    \includegraphics[scale=0.49]{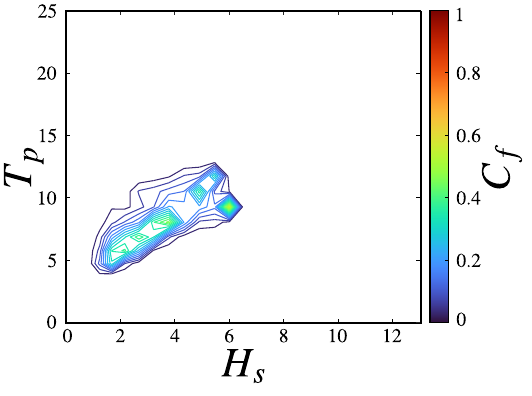}
    \caption{East Coast.}
    \label{subfig:EC_location_CF}
    \end{subfigure}%
    \captionsetup[figure]{justification=centering}
    \caption{Capacity factor of a single WEC with a prescribed design. The rated power is calculated as the maximum weighted power from the West Coast simulation and is used in the East Coast simulation (which results in a lower capacity factor and, thus, lower economic competitiveness of the WEC for the East Coast).}
    \label{fig:Site_Location}
\end{figure}

As an example, a WEC farm designed for a high-wave-energy region, such as the US West Coast or Alaska, but located in a low-wave-energy site, such as the US East Coast, might be capable of capturing a greater percentage of the available resource (compared to a WEC farm designed for a low-wave-energy site). However, the capacity factor, which is defined as the ratio of average power to the rated power, will be much lower.
This outcome can negatively affect the economic competitiveness of the wave farm, as the higher-rated power of the WEC is an over-design for the low-wave-energy resource site.
This is illustrated in Fig.~\ref{fig:Site_Location} for a single WEC with a fixed design.
Here, it is assumed that the maximum weighted power of the WEC from the West Coast is set as the rated power and is then used with the same WEC design in an East Coast simulation.
As the figure shows, the device's capacity factor on the East Coast is significantly lower, negatively affecting the WEC's economic competitiveness.

In addition to the implications of wave energy resource on capacity factor, the economic viability of the WEC farm can be affected through mooring costs, which tend to scale up with higher wave energy resource regions, often located in higher depths.
For example, according to Ref.~\cite{neary2014methodology}, mooring is expected to take up $8\%$ of the capital cost. 
In contrast, lower ocean depths have a significant impact on reducing the highest significant wave heights or reducing their probability \cite{MonarchaFernandes2013}, negatively limiting power generation opportunities. 
Consideration of such trade-offs (mooring requirements and power capture) informs design decisions that are tightly related to the selected site and its resource availability.

The dependency of WEC farm plant and layout decisions on the location of the site is investigated by implementing a concurrent plant and layout optimization problem for a 5-WEC farm, with prescribed controller parameters ($\BPTO = 500~[\unit{kNs/m}]$ and $\KPTO = -5~[\unit{kN/m}]$).
The exclusion of control variables from this concurrent optimization problem is motivated by some of the limitations associated with the control optimization problem, which were discussed in Sec.~\ref{sec:introduction}. 
The candidate sites are located on the West and East Coast, representing high and low energy resources, respectively.
The resulting optimization problem is highly non-linear, motivating the usage of a global optimizer, such as GA.
In spite of high computational costs, the results presented here are obtained using MS and are reported in Table~\ref{Tab:Study1}.
From this table, it is clear that the WEC radius increases by around $26\%$ for the East Coast region.
The slenderness ratio also increases by around $26\%$; however, the draft dimension remains at its lower bound of $0.5~[\unit{m}]$.
Although further investigations are required to understand the role of draft dimension on WEC farm performance, the draft value found in this solution seems to be related to the choice of the objective function.

Table~\ref{Tab:Study1} also offers some insights into the power captured by the farm, which is drastically higher for the high-energy-resource region.
Interactions among WEC devices are quantified using q-factor:
\begin{equation}
    \label{eq:q-factor}
    q_{\text{factor}} = \frac{\text{Power Captured by the Farm}}{\text{Device Power}\times \text{Number of WECs in the farm}}
\end{equation}

\noindent
The value of the q-factor is a function of WEC geometry, layout configuration, mooring, control strategy, and site location \cite{Goeteman2018}.  
It seems that while both solutions entail overall \textit{constructive} interactions (i.e., $q_{\text{factor}}$>1), the interaction effects for the East Coast region are better leveraged to improve the farm performance.
This result might be related to the performance of GA.
In that case, using a multi-start strategy may enable a solution with a higher q-factor on the West Coast.

\begin{figure}[t]
    \captionsetup[subfigure]{justification=centering}
    \centering
    \begin{subfigure}{\columnwidth}
    \centering
    \includegraphics[scale=0.5]{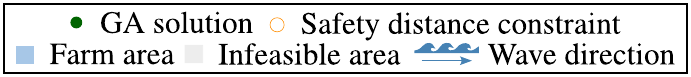}
    \label{subfig:legend}
    \end{subfigure}
    \begin{subfigure}{0.5\columnwidth}
    \centering
    \includegraphics[scale=0.8]{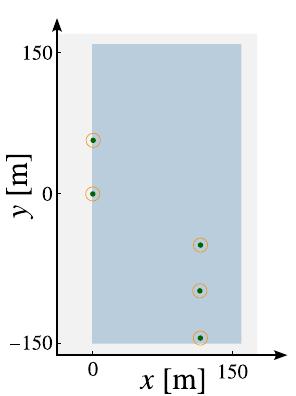}
    \caption{West Coast.}
    \label{subfig:LPWest}
    \end{subfigure}%
    \begin{subfigure}{0.5\columnwidth}
    \centering
    \includegraphics[scale=0.8]{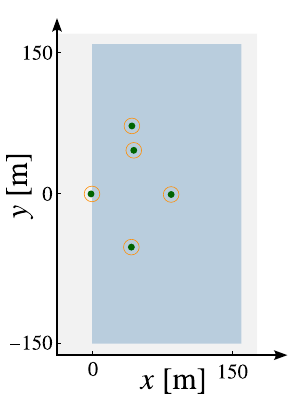}
    \caption{East Coast.}
    \label{subfig:LPEast}
    \end{subfigure}%
    \captionsetup[figure]{justification=centering}
    \caption{Optimized layout from the concurrent plant and layout optimization study for West and East costs. Green circles are drawn in scale based on the optimized radius of the WEC devices. The orange circles, also drawn in scale, indicate a safety distance constraint imposed to allow maintenance ships to pass.} 
    \label{fig:PLEastWestCoast}
\end{figure}

The optimized array for each site location is shown in Fig.~\ref{fig:PLEastWestCoast}.
The figure clearly shows that the high-energy resource area of the West Coast uses two clusters of WECs, each in the shape of a column.
This configuration seems to minimize the masking effect.
On the other hand, at the East Coast location, the optimized layout is configured to increase interactions among WEC devices.
From these results, it is evident that the geometrical and layout design of WEC farms are closely related to the candidate site.

\begin{table}[t]
\renewcommand{\arraystretch}{1.2}
  \begin{threeparttable}[t]
    \setlength{\abovecaptionskip}{0pt}
    \caption{Concurrent optimization of plant and layout using \texttt{MATLAB}'s genetic algorithm for West Coast and East Coast locations. The hydrodynamic coefficients are estimated through MS. The control parameters are prescribed as $\BPTO = 500~[\unit{kNs/m}]$, and $\KPTO = -5~[\unit{kN/m}]$, with no power saturation limits.}
  \begin{tabularx}{\linewidth}{@{}Y@{}}
   \label{Tab:Study1}
   \begin{tabular}{r r r r r r r r}
     \hline  \hline
    \multirow{2}{*}{\rotatebox[origin=c]{0}{\textbf{Site}}} &  \multicolumn{2}{c}{\textrm{\textbf{Plant}}} &  \multirow{2}{*}{\textrm{\textbf{Layout}}}& \multirow{2}{*}{\textrm{{P}}\tnote{b}} & \multirow{2}{*}{$q_{\text{factor}}$} \\ \cline{2-3}
     &  $\Rwec$\tnote{a} & $\ARwec$ &  &  &  \\
    \hline 
    \multirow{1}{*}{\rotatebox[origin=c]{0}{\textrm{West Coast}}} & $2.92$ & $5.84$ & \multirow{2}{*}{Fig.~\ref{fig:PLEastWestCoast}} & $71.5$ & $1.01$  \\  
    \multirow{1}{*}{\rotatebox[origin=c]{0}{\textrm{East Coast}}} & $3.68$ & $7.35$ & & $16.5$ & $1.05$ \\
    \hline \hline
    \end{tabular}
    \end{tabularx}
    \begin{tablenotes} [para,flushleft]
                \item [a] Calculated in [\unit{m}] \item [b] Calculated in [\unit{MW}]
     \end{tablenotes}
  \end{threeparttable}
\end{table}

Power extraction from a WEC device is largely affected by PTO damping and stiffness coefficients. 
The choice of these parameters depends to some extent on the location of the candidate site.
These variables also inform the PTO rating of the device and can have a significant impact on the electric and power electronic equipment required \cite{Tedeschi2012}.  
While control parameter optimization is best posed in the time domain, where constraints on device motion and control force can be imposed, for the sake of this discussion, we use frequency-domain models to exhibit the dependency of control parameters to site location.
To this end, we use various power saturation limits to bound the power capture in the WEC power matrix.
The results from this investigation are shown in Table.~\ref{tab:Control_site}.
While $\KPTO$ remains unchanged, optimal $\BPTO$ is different for each site location, and it varies for different power saturation limits.
Further investigations are required to understand the impact and behavior of $\KPTO$. Nevertheless, the results of this investigation point to the impact of site location on controller design.
Specifically, the value of $\BPTO$ is consistently smaller for the low-energy-resource region.

\begin{table}[h]
\renewcommand{\arraystretch}{1.1}
  \begin{threeparttable}[t]
  \setlength{\abovecaptionskip}{0pt}
      \caption{Optimized control parameters for various levels of power saturation limit for a single WEC device located on the West Coast and East Coast. \texttt{MATLAB}'s \texttt{fmincon} was used to solve the optimization problems.}
  \begin{tabularx}{\linewidth}{@{}Y@{}}
    \label{tab:Control_site}
       \centering
    \begin{tabular}{r r r r r r r}
        \hline \hline
        \multirow{2}{*}{\rotatebox[origin=c]{0}{\textbf{Site}}} & \multicolumn{2}{c}{\textbf{Plant}} & \multirow{2}{*}{\rotatebox[origin=c]{0}{$p_{\text{limit}}$}\tnote{b}} & \multicolumn{2}{c}{\textbf{Control}\tnote{c}} & \multirow{2}{*}{\textrm{{P}}\tnote{d}} \\
        \cline{2-3} \cline{5-6}
        & $\Rwec\tnote{a}$ & $\ARwec$ & & $\BPTO$ & $\KPTO$ & \\
        \hline 
        \multirow{5}{*}{\rotatebox[origin=c]{90}{West Coast}} &  \multirow{10}{*}{\rotatebox[origin=c]{0}{$5.00$}} &  \multirow{10}{*}{\rotatebox[origin=c]{0}{$5.00$}}& $50$ & $251.8$ & $-500.0$ & $0.9$ \\
         & & & $150$ & $253.4$ & $-500.0$ & $2.5$ \\
         & & & $250$ & $254.0$ & $-500.0$ & $4.0$ \\
         & & & $350$ & $218.3$ & $-500.0$ & $5.39$  \\
         & & & \text{None} & $308.7$ & $-500.0$ & $134.4$ \\ \cline{4-7}
        \multirow{5}{*}{\rotatebox[origin=c]{90}{East Coast}} &   & & $50$ & $165.9$ & $-500.0$ & $0.4$ \\
         & & & $150$ & $165.6$ & $-500.0$ & $1.0$ \\
         & & & $250$ & $157.6$ & $-500.0$ & $1.4$ \\
         & & & $350$ & $146.2$ & $-500.0$  & $1.8$ \\
         & & & \text{None} & $131.26$ & $-500.0$ & $19.6$ \\
        \hline \hline
    \end{tabular}
        \end{tabularx}
        \begin{tablenotes} [para,flushleft]
                \item [a] Prescribed in [\unit{m}] \item [b] Prescribed in [\unit{kW}]  \item [c] Calculated in [\unit{kNs/m}] and [\unit{kN/m}] \item [d] Calculated in [\unit{MW}]
     \end{tablenotes}
  \end{threeparttable}
\end{table}

%----------------
% P_wave

\begin{postulate}\label{P_Wave}
Optimal WEC farm design is sensitive to the wave type.
\end{postulate}
It is often convenient to investigate the design of a WEC farm for a single or a few regular or harmonic waves. 
Real sea states, however, are composed of several wave trains, each mathematically represented as the superposition of several harmonic waves, each with a specific height and wave period.
When it is assumed that these waves travel in the same direction, they are referred to as long-crested irregular waves \cite{Jacobson2011}.
Using long-crested irregular waves in the design of a WEC farm often results in a smoothing effect on array interactions.
This effect, which is due to the presence of waves with a multiple of frequencies, reduces the intensity of array interactions compared to when regular waves are considered \cite{Han2023, Balitsky2018}.
The importance of using irregular waves is further discussed in detail in Ref.~\cite{RASHIDI2024131717}.

\begin{figure}[t]
    \captionsetup[subfigure]{justification=centering}
    \centering
    \begin{subfigure}{0.48\columnwidth}
    \centering
    \includegraphics[scale=0.50]{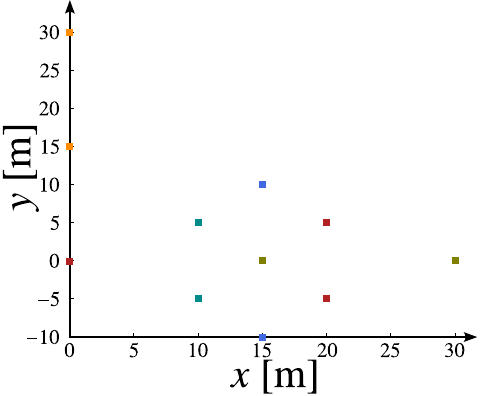}
    \caption{Five prescribed farm layouts.}
    \label{subfig:WC_location}
    \end{subfigure}%
    \begin{subfigure}{0.48\columnwidth}
    \centering
    \includegraphics[scale=0.50]{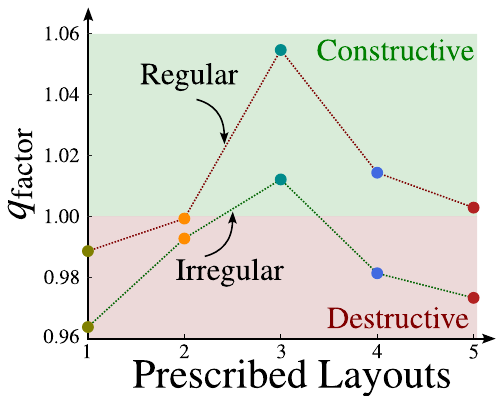}
    \caption{$q_\text{factor}$}
    \label{subfig:Qfactor_5layouts}
    \end{subfigure}%
    \captionsetup[figure]{justification=centering}
    \caption{Smoothing effect appearing in the q-factor calculation for a range of WEC farm layouts for unidirectional irregular waves.}
    \label{fig:Wave_rVir_Q}
\end{figure}

To assess whether the interaction effects within a farm are constructive or destructive, q-factor is used.
To visualize the smoothing effect that results from long-crested irregular waves, we first prescribed five different farm layouts, each with $3$ WEC devices (see Fig.~\ref{subfig:WC_location}).
The prescribed layouts for these studies constitute cases with \textit{close} proximity of WECs.
Using a non-optimal controller, the simulation studies were carried out for both regular and irregular waves, as shown in Fig.~\ref{subfig:Qfactor_5layouts}. 
These figures clearly show that the farm's q-factor is drastically reduced for unidirectional irregular waves.
Note that as the distance between WEC devices increases, interactions among them are reduced, and the q-factor approaches one.
For regular waves, it is reported that even at $2000~[\unit{m}]$ separation distance, up to $15\%$ interactions can occur.
This, however, reduces to less than $10\%$ for distances greater than $400~[\unit{m}]$ for irregular waves \cite{Babarit2010}.

Short-crested (multi-directional) irregular waves present a more realistic approach to modeling the wave climate.
It is known that the q-factor estimated using unidirectional waves is often an overestimate \cite{Han2023a}.
This outcome is because high values of the q-factor for a specific wave direction are associated with poor q-factors at some other directions \cite{GarciaRosa2015}.
This statement begs more questions regarding the impact that integrated design studies, such as concurrent plant, control, and layout optimization, can have on the performance of the WEC farm.
Particularly, considering this smoothing effect, it is unclear when including layout optimization into the design and control optimization of WEC farms would be justified.
This area is one of the future research directions for the WEC design community. 

Considering the variability in ocean waves, it is important to not only use irregular waves but also utilize the joint probability distribution of the waves.
This recommendation is because WEC devices designed for the largest waves often exhibit low capacity factors \cite{Coe2021}.  
This consideration brings in impacts from the site location since the probability distribution of waves is strongly dependent on the site of interest. 
The impact of irregular waves on WEC farm design is clear due to the fact that real sea states entail components with various frequencies, occurring with a probability that can be theoretically estimated.
Designing the WEC farm based on isolated regular waves, therefore, is far from realistic.

\begin{figure}[t]
    \captionsetup[subfigure]{justification=centering}
    \centering
    \begin{subfigure}{0.49\columnwidth}
    \centering
    \includegraphics[scale=0.50]{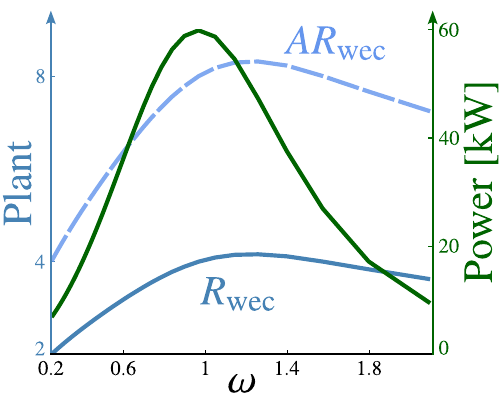}
    \caption{Optimal plant.}
    \label{subfig:Plant_reg_wavetype}
    \end{subfigure}%
    \begin{subfigure}{0.49\columnwidth}
    \centering
    \includegraphics[scale=0.50]{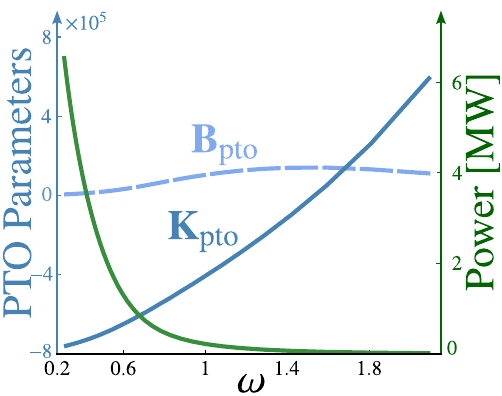}
    \caption{Optimal control.}
    \label{subfig:Control_reg_wavetype}
    \end{subfigure}%
    \captionsetup[figure]{justification=centering}
    \caption{Optimal plant and control variables for regular waves.}
    \label{fig:Wave_regular_OPts}
\end{figure}

The sensitivity of optimal plant and control to the wave type becomes clear once we recognize the fact that these values change for different frequencies of regular waves.
This statement is illustrated in Figs.~\ref{subfig:Plant_reg_wavetype}, and \ref{subfig:Control_reg_wavetype} for optimal plant and control variables, respectively.
Since irregular waves entail a multitude of frequencies, the optimal plant and control solutions for irregular waves are often related to dominant wave frequencies and their probability of occurrence.

\begin{figure}[t]
    \captionsetup[subfigure]{justification=centering}
    \centering
    \begin{subfigure}{0.49\columnwidth}
    \centering
    \includegraphics[scale=0.50]{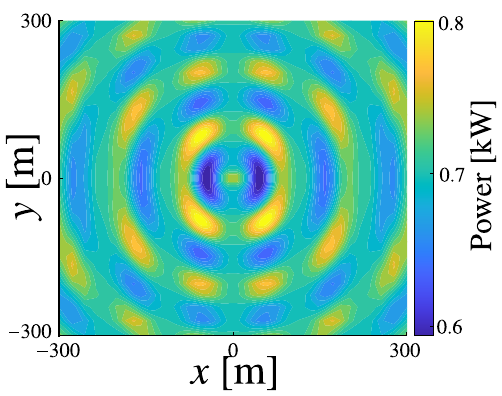}
    \caption{Regular Wave type.}
    \label{subfig:layour_reg_wavetype}
    \end{subfigure}%
    \begin{subfigure}{0.49\columnwidth}
    \centering
    \includegraphics[scale=0.50]{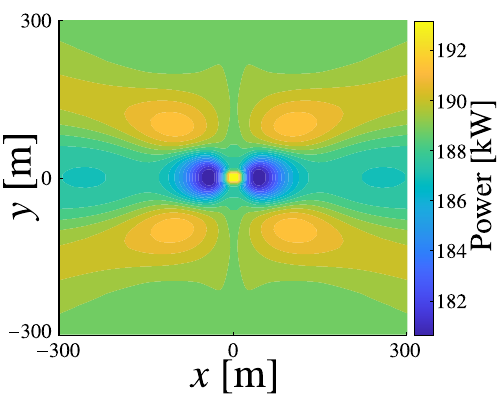}
    \caption{Irregular wave type.}
    \label{subfig:layour_irreg_wavetype}
    \end{subfigure}%
    \captionsetup[figure]{justification=centering}
    \caption{Power landscape in terms of layout for regular and irregular wave types for two WECs with prescribed design and control parameters.}
    \label{fig:Wavetype_layout}
\end{figure}

To highlight the impact of wave type on the layout design of the farm, we first prescribed the plant and control parameters of the WEC device. 
Next, we place the first WEC at the center of the coordinate system while moving the second one in the layout space for both regular and irregular waves.
These series of simulations enable the visualization of the layout space in terms of power, making direct comparisons between layouts in regular and irregular waves simpler.
These simulations are shown in Fig.~\ref{subfig:layour_reg_wavetype} and \ref{subfig:layour_irreg_wavetype} for regular and irregular waves, respectively.
From these figures, it is clear that when irregular waves are considered, the layout and, thus, its optimal configuration are different from regular cases.

%----------------
% P_Plant

\begin{postulate}\label{P_Plant}
Optimal WEC plant design is sensitive to variations in control, layout, and site location. 
\end{postulate}

The geometry of the WEC device directly affects its performance and cost.
Larger WEC devices are often capable of capturing higher amounts of power at a smaller range of frequencies, while
smaller devices capture smaller amounts of power at a wider range of frequencies \cite{Coe2021}.
Therefore, it is important to optimally size the geometry of the WEC in the farm by considering key design drivers, such as control, layout, and the selected site. 
While a power per unit volume objective function \cite{falnes2020ocean} is a reasonable choice, an economic objective function, such as levelized cost of energy (LCOE), might be better capable of capturing the intricate relationship between device dimensions, and its manufacturing, operation, and maintenance costs.

In this section, our objective is to investigate the potential coupling between optimal plant design and PTO control, layout considerations, and site selection.
Note that while plant coupling with site selection was established in Postulate~\ref{P_site}, additional results based on various power saturation limits are provided in this section.  
Mathematically, this coupling can be described using the following equations:
\begin{equation}
\label{eq:plantcoupling}
    \frac{\partial \bm{p}^{*}}{\partial\bm{u}} \neq \bm{0}, ~~ \frac{\partial \bm{p}^{*}}{\partial\AL} \neq \bm{0}, ~~
    \frac{\partial \bm{p}^{*}}{\partial \bm{L}} \neq \bm{0}
\end{equation}

\noindent
where $\parm^{*}$ refers to the optimal solution.
The first term in Eq.~(\ref{eq:plantcoupling}) should be interpreted as if we modify the control parameters, the \textit{optimal} solution for the plant will be different.
Note that these equations should be placed within the broader scope of the optimization problem formulated in Eq.~(\ref{Eqn:OPtimization}) because problem constraints can have a large impact on the interpretation of the presented results. 

To demonstrate plant coupling with control, we use various power saturation limits with fixed control parameters to solve a series of plant optimization problems.
The optimal $\Rwec$ and $\ARwec$ are reported in Table~\ref{tab:P3:plant_on_control} for a high-energy-resource region on the West Coast and low-energy-resource on the East Coast. 
\begin{table}[t]
\renewcommand{\arraystretch}{1.2}
  \begin{threeparttable}[t]
      \setlength{\abovecaptionskip}{0pt}
      \caption{Optimized plant with prescribed PTO parameters and different power saturation limits for a single WEC device located on the West Coast and East Coast. The hydrodynamic coefficients were calculated using MS, and the optimization problem was solved using \texttt{fmincon}.}
  \begin{tabularx}{\linewidth}{@{}Y@{}}
    \label{tab:P3:plant_on_control}
       \centering
    \begin{tabular}{r r r r r r r}
        \hline \hline
        \multirow{2}{*}{\rotatebox[origin=c]{0}{\textbf{Site}}} & \multicolumn{2}{c}{\textbf{Control}\tnote{a}} & \multirow{2}{*}{\rotatebox[origin=c]{0}{$p_{\text{limit}}\tnote{b}$}} & \multicolumn{2}{c}{\textbf{Plant}} & \multirow{2}{*}{\textrm{{P}}\tnote{c}}\\
        \cline{2-3} \cline{5-6}
        & $\KPTO$ & $\BPTO$ & & $R_{WEC}~[\unit{m}]$ & $AR_{WEC}$ \\
        \hline 
        \multirow{4}{*}{\rotatebox[origin=c]{90}{West Coast}} &  \multirow{8}{*}{\rotatebox[origin=c]{0}{$-0.5$}} &  \multirow{8}{*}{\rotatebox[origin=c]{0}{$500$}}& $1$ & $0.5$ & $1$ & $0.01$ \\
        & & & $10^2$ & $1.23$ & $2.46$ & $0.48$ \\
        & & & $10^5$ & $2.91$ & $5.81$ & $13.82$ \\
        & & & None & $2.91$ & $5.81$ & $13.82$  \\
        \cline{4-7}
        \multirow{4}{*}{\rotatebox[origin=c]{90}{East Coast}} &   &  & $1$ & $0.63$ & $1.26$ & $0.01$ \\
        & & & $10^2$ & $1.72$ & $3.44$ & $0.18$ \\
        & & & $10^5$ & $3.56$ & $7.12$ & $2.92$ \\
        & & & None & $3.56$ & $7.12$ & $2.92$ \\
        \hline \hline
    \end{tabular}
        \end{tabularx}
        \begin{tablenotes} [para,flushleft]
       \item [a] Calculated in [\unit{kNs/m}] and [\unit{kN/m}] \item [b] Prescribed in [\unit{kW}] \item [c] Calculated in [\unit{MW}]
     \end{tablenotes}
  \end{threeparttable}
\end{table}

From this table, it is clear that $\Rwec$ and $\ARwec$ change for various power saturation limits, for each site location, until they plateau for the case with no power limit imposed.
This result indicates differences in the overall power absorption capabilities of the WEC device at each location.  
The low-energy resource demands a WEC with a larger radius.
This can be related to the probability distribution of waves (defined as a function of wave peak period and significant wave height) on the East Coast location.
For example, if the annual joint probability distribution of waves for the majority of the farm life is closely packed over a smaller range of frequencies, then using smaller devices that capture power at a lower amplitude from a wider range of frequencies may not be intuitive for this site.
On the other hand, if the West Coast location offers a wider range of frequencies for the majority of the years considered, then a smaller device may be more suitable for this location.

It is also evident that the WEC geometry varies for different power saturation limits, indicating that optimal WEC geometry is affected by the choice of PTO control.  
These results (interactions between control and optimal WEC geometry) are also corroborated by the WEC literature.
For example, Ref.~\cite{GarciaRosa2016} offers a detailed analysis of optimal WEC geometry sensitivity to latching, declutching, and model predictive control technologies.
Similarly, Ref.~\cite{Sjoekvist2014} discusses the impact of plant optimality with fixed controller designs on the performance of the device.

\begin{figure}[t]
    \captionsetup[subfigure]{justification=centering}
    \centering
    \begin{subfigure}{0.49\columnwidth}
    \centering
    \includegraphics[scale=0.50]{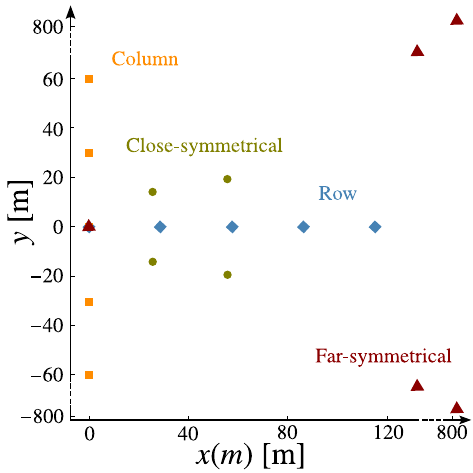}
    \caption{Layout configurations.}
    \label{subfig:Prescribedlayouts_plantlayout}
    \end{subfigure}%
    \begin{subfigure}{0.49\columnwidth}
    \centering
    \includegraphics[scale=0.50]{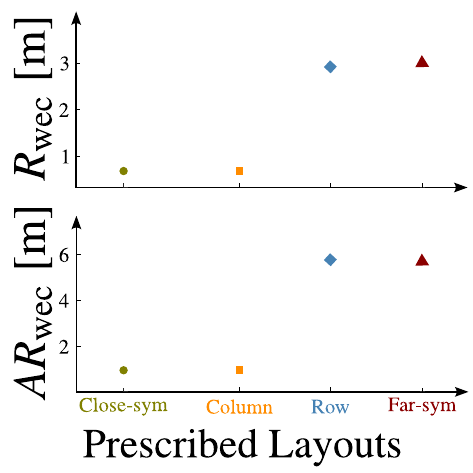}
    \caption{Optimal plant}
    \label{subfig:plantlayout}
    \end{subfigure}%
    \captionsetup[figure]{justification=centering}
    \caption{Optimal plant solution for four different layout configurations.}
    \label{fig:PLant_layout}
\end{figure}

Device geometry has a considerable impact on array interactions. 
According to Ref.~\cite{Han2023}, flat truncated cylinders produce more constructive interactions than slender cylinders. 
The dependency of optimal plant on the layout of the WEC farm can be established by solving plant optimization problems for different layout configurations.
To this end, we use the layouts presented in Fig.~\ref{subfig:Prescribedlayouts_plantlayout} and solve a series of plant-only optimization problems for a $5$-WEC farm located on the Alaska Coast.
PTO parameters are prescribed as $\BPTO = 100~[\unit{kNs/m}]$, and $\KPTO = -500 ~[\unit{kN/m}]$.

The optimal plant solutions are presented in Fig.~\ref{subfig:plantlayout}.
From the figure it is clear that the optimal radius is $0.5~[\unit{m}]$ for close symmetrical, and column layout configurations, but it changes to $2.93~[\unit{m}]$, and $3.01~[\unit{m}]$ for row, and far symmetrical configurations, respectively.
The slenderness ratio also changes for these layouts, indicating that a coupling between layout and optimal plant exists.
This coupling is further explored by conducting concurrent layout and dimension optimization studies in Refs.~\cite{Abdulkadir2023, Lyu2019, Azad24}.

%---------------
% P_control

\begin{postulate}\label{P_Control}
Optimal WEC control design is sensitive to variations in plant, layout, and site selection.
\end{postulate}

\begin{table}[t]
\renewcommand{\arraystretch}{1.2}
  \begin{threeparttable}[t]
  \setlength{\abovecaptionskip}{0pt}
      \caption{Optimized control for different power saturation limits, with a fixed layout configuration (close-symmetrical, as shown in Fig.~\ref{subfig:Prescribedlayouts_plantlayout}), located on the West Coast. The hydrodynamic coefficients were calculated using multiple scattering, and the optimization problem was solved using \texttt{fmincon}.}
  \begin{tabularx}{\linewidth}{@{}Y@{}}
  \label{tab:P3:control_on_plant}
       \centering
    \begin{tabular}{r r r r r r r r}
        \hline \hline
         \multirow{2}{*}{\textbf{Site}} & \multicolumn{2}{c}{\textbf{Plant}} &  \multirow{2}{*}{\rotatebox[origin=c]{0}{$p_{\text{limit}}$\tnote{b}}} & \multicolumn{2}{c}{\textbf{Control}\tnote{c}} & \multirow{2}{*}{\textrm{{P}}\tnote{d}} \\ \cline{2-3} \cline{5-6}
        & $\Rwec$\tnote{a} & $\ARwec$ & & $\BPTO$ & $\KPTO$ & \\
        \hline
        \multirow{8}{*}{\rotatebox[origin=c]{90}{West Coast}} & \multirow{4}{*}{$5$} & \multirow{4}{*}{$5$} & $1$ & $357.1$ & $-500.0$  & $0.1$ \\
        & & & $10$ & $198.3$ & $-500.0$ & $1.0$\\
        & & & $100$ & $313.0$ & $-500.0$ & $8.7$ \\
        & & & \text{None} & $372.4$ & $-500.0$ & $530.9$ \\ \cline{2-7}
        & \multirow{4}{*}{$10$} & \multirow{4}{*}{$5$}  & $1$ & $500.0$ &  $-500.0$ & $0.1$ \\
        & & & $10$ & $500.0$ & $-500.0$ & $1.0$ \\
        & & & $100$ & $500.0$ & $-500.0$ & $7.9$ \\
        & & & \text{None} & $500.0$ & $-500.0$ & $207.5$ \\
        \hline \hline
    \end{tabular}
        \end{tabularx}
        \begin{tablenotes} [para,flushleft]
       \item [a] Calculated in [\unit{m}] \item [b] Prescribed in ~[\unit{kW}] \item [c] Calculated in [\unit{kNs/m}] and [\unit{kN/m}] \item [d] Calculated in [\unit{MW}]
     \end{tablenotes}
  \end{threeparttable}
\end{table}

The control technology deployed in WEC devices has a large impact on power capture, as well as its operation.
While an aggressive controller may increase power capture through large motions/forces, it has the potential to incur damages and negatively affect the economic competitiveness of the farm \cite{Ringwood2023}.
Therefore, optimizing the WEC controller in the presence of practical PTO limits and motion constraints is a necessary step in improving the performance of these devices.

Under the assumptions discussed in Sec.~\ref{sec:Background}, this section demonstrates the dependency of optimal WEC control on plant, layout, and site selection. 
Mathematically, this can be described as:
\begin{equation}
\label{eq:controlcoupling}
    \frac{\partial \bm{u}^{*}}{\partial\bm{p}} \neq \bm{0}, ~~ \frac{\partial \bm{u}^{*}}{\partial\AL} \neq \bm{0}, ~~
    \frac{\partial \bm{u}^{*}}{\partial \bm{L}} \neq \bm{0}
\end{equation}

\noindent
Note that here, instead of imposing direct constraints on device motion or PTO force, the power matrix of WEC devices is saturated at prescribed limits.
More advanced implementations of control-related constraints are studied in the literature.
For example, Ref.~\cite{PenaSanchez2022} studied the impact of PTO amplitude and force constraints on the performance of a WEC system in the context of control co-design.

The coupling between control and plant domains is established by solving a series of control optimization problems for a $5$-WEC farm.
The layout of the farm is fixed in the close-symmetrical configuration, shown in Fig.~\ref{subfig:Prescribedlayouts_plantlayout}.
The radius of the WEC device is changed from $\Rwec = 5~[\unit{m}]$ to $10~[\unit{m}]$, with the slenderness ratio kept fixed at $\ARwec = 5$.
The second specification results in a more slender device.
The results, which are presented in Table~\ref{tab:P3:control_on_plant}, indicate that for a given power saturation limit, optimal $\BPTO$ changes, while $\KPTO$ remains unchanged.
While these results show the optimal control solution's sensitivity to plant variations, they also highlight the impact of practical PTO force considerations.
In addition, by comparing the power that was captured by the farm, it is evident that slender devices have a relatively lower performance.
These observations, however, require further investigations.

While the dependency of optimal control to selected site was established in Postulate~\ref{P_site}, further results for this coupling are provided in Table~\ref{tab:P3:control_on_site}, by solving the optimization problems from Table~\ref{tab:P3:control_on_plant}, with the fixed plant design of $\Rwec = 5~[\unit{m}]$ and $\ARwec = 5$ for the East Coast. 
When compared to Table~\ref{tab:P3:control_on_plant}, the results indicate that optimal control parameters are sensitive to site location.
Here, the optimal $\BPTO$ has also comparatively reduced for the low-energy-resource location.

\begin{table}[t]
\renewcommand{\arraystretch}{1.2}
  \begin{threeparttable}[t]
   \setlength{\abovecaptionskip}{0pt}
  \caption{Optimized control for different power saturation limits, with a fixed layout configuration (close-symmetrical, as shown in Fig.~\ref{subfig:Prescribedlayouts_plantlayout}), located on the East Coast. The hydrodynamic coefficients were calculated using MS, and the optimization problem was solved using \texttt{fmincon}.}
  \begin{tabularx}{\linewidth}{@{}Y@{}}
    \label{tab:P3:control_on_site}
       \centering
    \begin{tabular}{r r r r r r r r}
        \hline \hline
        \multirow{2}{*}{\rotatebox[origin=c]{0}{\textbf{Site}}} & \multicolumn{2}{c}{\textbf{Plant}} &  \multirow{2}{*}{\rotatebox[origin=c]{0}{$p_{\text{limit}}\tnote{b}$}} & \multicolumn{2}{c}{\textbf{Control}\tnote{c}} & \multirow{2}{*}{\textrm{{P}}\tnote{d}} \\ \cline{2-3} \cline{5-6}
        & $\Rwec$\tnote{a} & $\ARwec$ & & $\BPTO$ & $\KPTO$  & \\ 
        \hline
        \multirow{4}{*}{\rotatebox[origin=c]{90}{\text{East Coast}}} & \multirow{4}{*}{\rotatebox[origin=c]{0}{$5$}} & \multirow{4}{*}{\rotatebox[origin=c]{0}{$5$}} & $1$ & $206.9$ & $-288.0$ & $0.05$\\
         & & & $10$ & $115.3$ & $-449.8$ & $0.49$ \\
         & &  & $100$ & $134.7$ & $-500.0$ & $3.65$ \\ 
         & & & \text{None} & $128.9$ & $-497.9$ & $30.42$ \\
        \hline \hline
    \end{tabular}
        \end{tabularx}
        \begin{tablenotes} [para,flushleft]
        \item [a] Calculated in [\unit{m}] \item [b] Prescribed in [\unit{kW}] \item [c] Calculated in [\unit{kNs/m}] and [\unit{kN/m}] \item [d] Calculated in [\unit{MW}]
     \end{tablenotes}
  \end{threeparttable}
\end{table}

The dependency of the optimal control to layout configuration is investigated by solving a series of optimal control problems using MS, for row and close-symmetrical layout configurations, shown in Fig~\ref{subfig:Prescribedlayouts_plantlayout}.
The results from this investigation are reported in Table~\ref{tab:P3:control_on_Layout}.

\begin{table}[t]
\renewcommand{\arraystretch}{1.2}
  \begin{threeparttable}[b]
   \setlength{\abovecaptionskip}{0pt}
      \caption{Optimized control for different power saturation limits, for row and close-symmetrical layout configurations of a $5$-WEC farm located on the West Coast. The hydrodynamic coefficients were calculated using multiple scattering, and the optimization problem was solved using \texttt{fmincon}.}
  \begin{tabularx}{\linewidth}{@{}Y@{}}
    \label{tab:P3:control_on_Layout}
       \centering
    \begin{tabular}{r r r r r r r r}
        \hline \hline
        \multirow{2}{*}{\rotatebox[origin=c]{0}{\textbf{Layout}}} & \multicolumn{2}{c}{\textbf{Plant}} &  \multirow{2}{*}{\rotatebox[origin=c]{0}{$p_{\text{limit}}$\tnote{a}}} & \multicolumn{2}{c}{\textbf{Control}\tnote{b}} & \multirow{2}{*}{\textrm{{P}}\tnote{c}} \\ \cline{2-3} \cline{5-6} 
        & $\Rwec$ & $\ARwec$ & & $\BPTO$ & $\KPTO$ & \\
        \hline
        \multirow{4}{*}{\rotatebox[origin=c]{90}{\text{Close-sym}}} & \multirow{8}{*}{\rotatebox[origin=c]{0}{$5$}} & \multirow{8}{*}{\rotatebox[origin=c]{0}{$5$}} & $1$ & $357.1$ & $-500$ & $0.1$ \\
         & & & $10$ & $198.3$ & $-500$ & $1.0$  \\
         & &  & $100$ & $313.0$ & $-500$ & $8.7$ \\ 
         & & & \text{None} & $372.4$ & $-500$ & $530.9$ \\ \cline{4-7}
                 \multirow{4}{*}{\rotatebox[origin=c]{90}{\text{Row}}} & &  & $1$ & $350.7$ & $-500$ & $0.1$ \\
         & & & $10$ & $215.6$ & $-500$ & $1.0$  \\
         & &  & $100$ & $307.2$ & $-500$ & $8.7$ \\ 
         & & & \text{None} & $355.7$ & $-500$ & $553.72$  \\
        \hline \hline
    \end{tabular}
        \end{tabularx}
        \begin{tablenotes} [para,flushleft]
        \item [a] Prescribed in [\unit{kW}] \item [b] Calculated in [\unit{kNs/m}] and [\unit{kN/m}] \item [c] Calculated in [\unit{MW}]
     \end{tablenotes}
  \end{threeparttable}
\end{table}

From this figure, it is clear that PTO parameters change for different layout configurations, and power saturation limits.
The power captured by the farm remains very similar for both layout configurations when power saturation limits are imposed.
This points to some of the limitations of directly saturating the WEC power matrix, because this approach may remove contributions from neighboring devices.  
Note also that when no power constraint is imposed, the row layout is capable of capturing more power than the close-symmetrical array.
This observation may be associated with the optimizer finding a better control solution for the row layout.
It may also be associated with the WEC device interactions within the farm.
Specifically, the q-factors for the arbitrarily prescribed row and close-symmetrical layouts are $0.83$, and $0.79$, respectively.

These results are consistent with our expectations.
Specifically, it is known that an optimal controller for the heaving cylinder WEC device is often a resonator, resulting in large amplitudes of the device motion.
Therefore, the choice of control parameters, and its limits can have a major impact on device motion, and in turn, on the performance of all surrounding devices.
This impact becomes evident once we consider the equation of WEC natural frequency \cite{Ning2022}:
\begin{equation}
    \label{eq:Resonance}
    \omega_n = \sqrt{\frac{\bm{k}_{\text{pto}} + G}{\mathbf{M} + \mathbf{A}(\omega,\mathbf{w})}}
\end{equation}
\noindent
where $G$ is the hydrostatic coefficient, $\mathbf{M}$ is the mass, and $\mathbf{A}(\cdot)$ is the added mass matrix, which entails contributions from the surrounding devices.
Note that the added mass matrix, which is a function of the farm layout, and WEC geometry, affects the natural frequency of the WEC devices.

The impact of the layout on optimal PTO control can be extended to make additional conclusions regarding control optimization of WECs.
For example, this dependency indicates that the optimal PTO control solution is different between a farm and a single device. 
In other words, when a controller is optimized for a single device (or farm), but used in the farm (or single device), the value of the q-factor, described in Eq.~(\ref{eq:q-factor}), is often reduced (increased), because the controller is off resonance for the latter case.
While there is a need for the quantification of this impact, this observation motivates the control optimization of WEC devices within a farm, when more integrated approaches are not employed.

This observation also highlights some of the limitations of q-factor in capturing the interaction effect when an integrated design approach with control optimization (such as CCD) is used.
Note that q-factor is designed to assess the interaction effects within a farm.
This purpose is well served through Eq.~(\ref{eq:q-factor}) when a plant, and/or layout optimization is studied. 
However, when control optimization is also involved, the optimal controller will be off-resonance for the single device.
Therefore, the value reported by the q-factor in this case, takes into consideration the impacts from interactions, as well as control optimality.

%---------------
% P_layout
\begin{postulate}\label{P_Layout}
Optimal WEC layout design is sensitive to plant, control, and the selected site. 
\end{postulate}

Layout optimization of WEC devices within a farm is necessitated by the fact that when placed in close proximity, WEC devices interact with each other through the scattering and radiation waves \cite{Babarit2013}.
In this section, we assess the sensitivity of optimal layout to plant, control, and the selected site.
Mathematically, this can be shown as:
\begin{equation}
\label{eq:layoutcoupling}
    \frac{\partial \bm{\AL}^{*}}{\partial\bm{p}} \neq \bm{0}, ~~ \frac{\partial \bm{\AL}^{*}}{\partial \bm{u}} \neq \bm{0}, ~~
    \frac{\partial \bm{\AL}^{*}}{\partial \bm{L}} \neq \bm{0}
\end{equation}

\noindent
Note that according to Eq.~(\ref{Eqn:OPtimization}), the optimal layout sensitivity is assessed in the presence of safety constraints that practically impose a minimum of $10~[\unit{m}]$ distance between WECs. 

\begin{figure*}[t]
    \captionsetup[subfigure]{justification=centering}
    \centering
    \begin{subfigure}{\columnwidth}
    \centering
    \includegraphics[scale=0.7]{Legend.pdf}
    \label{subfig:Legend}
    \end{subfigure}
    \begin{subfigure}{0.5\columnwidth}
    \centering
    \includegraphics[scale=0.9]{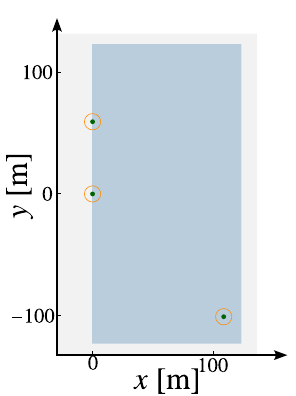}
    \caption{$\bm{p} = [2, 1]^{T}$.}
    \label{subfig:A_west_R05}
    \end{subfigure}%
    \begin{subfigure}{0.5\columnwidth}
    \centering
    \includegraphics[scale=0.9]{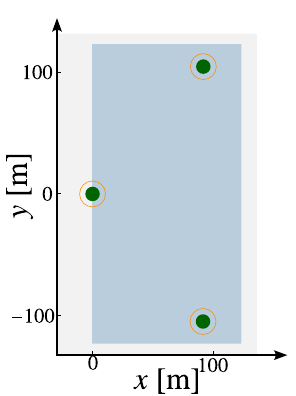}
    \caption{$\bm{p} = [6, 3]^{T}$.}
    \label{subfig:A_west_R8}
    \end{subfigure}%
    \begin{subfigure}{0.5\columnwidth}
    \centering
    \includegraphics[scale=0.9]{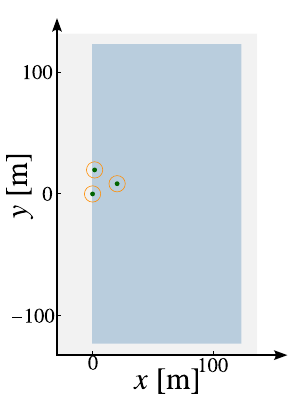}
    \caption{$\bm{u} = [2, -2]^{T}$.}
    \label{subfig:A_west_ctrl}
    \end{subfigure}%
    \begin{subfigure}{0.5\columnwidth}
    \centering
    \includegraphics[scale=0.9]{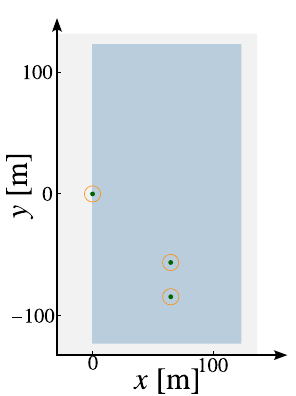}
    \caption{East Coast.}
    \label{subfig:A_east}
    \end{subfigure}%
    \captionsetup[figure]{justification=centering}
    \caption{Optimized layout for the demonstration of optimal layout sensitivity to plant, control, and site selection with WEC size and distance constraints drawn in scale in green and orange, respectively. Details of each study are presented in Table~\ref{tab:P3:Layout}.}
    \label{fig:Array} 
\end{figure*}

To assess the sensitivity of layout to the WEC plant, two layout optimization studies were implemented with fixed PTO parameters ($\BPTO = 500~[\unit{kNs/m}]$ and $\KPTO = -5~[\unit{kN/m}]$) on the West Coast.
The prescribed plant parameters are $\bm{p} = [2, 1]^{T}$, and $\bm{p} = [6, 3]^{T}$.
Using GA with MS for a $5$-WEC farm is computationally expensive.
Therefore, for the purposes of this discussion, we reduce the farm size to include $3$ WECs.
The problem settings and power captured by the farm for these studies are presented in Table~\ref{tab:P3:Layout}.
The optimized layout configurations are presented in Fig.~\ref{fig:Array}.

\begin{table}[t]
\renewcommand{\arraystretch}{1.2}
  \begin{threeparttable}[b]
  \setlength{\abovecaptionskip}{0pt}
      \caption{Optimized layout for a $3$-WEC farm located on the West Coast and East Coast, with prescribed plant and PTO parameters. The hydrodynamic coefficients were calculated using MS, and the optimization problem was solved using GA.}
  \begin{tabularx}{\linewidth}{@{}Y@{}}
    \label{tab:P3:Layout}
       \centering
    \begin{tabular}{r r r r r r r r}
        \hline \hline
        \multirow{2}{*}{\rotatebox[origin=c]{0}{\textbf{Site}}} & \multicolumn{2}{c}{\textbf{Plant}} & \multicolumn{2}{c}{\textbf{Control}\tnote{b}} & \multirow{2}{*}{\rotatebox[origin=c]{0}{\textbf{Layout}}} & \multirow{2}{*}{\textrm{{P}}\tnote{c}} \\ \cline{2-3} \cline{4-5} 
        & $\Rwec$\tnote{a} & $\ARwec$ & $\BPTO$ & $\KPTO$ &  & \\
        \hline
        \multirow{1}{*}{\rotatebox[origin=c]{0}{\text{West C.}}} & \multirow{1}{*}{\rotatebox[origin=c]{0}{$2$}} & \multirow{1}{*}{\rotatebox[origin=c]{0}{$1$}} & $500$ & $-5$ & Fig.~\ref{subfig:A_west_R05} & $15.28$ \\
        \multirow{1}{*}{\rotatebox[origin=c]{0}{\text{West C.}}} & \multirow{1}{*}{\rotatebox[origin=c]{0}{$6$}} & \multirow{1}{*}{\rotatebox[origin=c]{0}{$3$}} & $500$ & $-5$ & Fig.~\ref{subfig:A_west_R8} & $99.41$\\
        \multirow{1}{*}{\rotatebox[origin=c]{0}{\text{West C.}}} & \multirow{1}{*}{\rotatebox[origin=c]{0}{$2$}} & \multirow{1}{*}{\rotatebox[origin=c]{0}{$1$}} & $2$ & $-2$ & Fig.~\ref{subfig:A_west_ctrl} & $0.59$ \\
        \multirow{1}{*}{\rotatebox[origin=c]{0}{\text{East C.}}} & \multirow{1}{*}{\rotatebox[origin=c]{0}{$2$}} & \multirow{1}{*}{\rotatebox[origin=c]{0}{$1$}} & $500$ & $-5$ & Fig.~\ref{subfig:A_east} & $1.56$ \\
        \hline \hline
    \end{tabular}
        \end{tabularx}
        \begin{tablenotes} [para,flushleft]
        \item [a] Calculated in [\unit{m}] \item [b] Calculated in [\unit{kNs/m}] and [\unit{kN/m}] \item [c] Calculated in [\unit{MW}] 
     \end{tablenotes}
  \end{threeparttable}
\end{table}

According to Table~\ref{tab:P3:Layout}, when the WEC device geometry is changed, the power captured by the farm changes dramatically.
This result is directly related to the fact that larger WEC devices capture more power.
The optimized layouts associated with each of these cases are presented in Fig.~\ref{subfig:A_west_R05} and \ref{subfig:A_west_R8}.
In these figures, the WEC dimension and the associated constraints are drawn in scale.
Since the optimized layout is different for each case, we can conclude that, under the given assumptions, the optimal layout is sensitive to plant design.
The results seem to be in conflict with those offered in Ref.~\cite{Han2023}, where the authors concluded that the draft, slenderness ratio, and bottom shape of the buoy barely affect the q-factor and that a sequential approach, in which the optimization of the buoy geometry is followed by layout optimization would be sufficient.
This discrepancy in the conclusions begs further inspection. 
Specifically, our investigations indicate that contributions from layout configuration can be small for non-optimal controllers.
However, when the control optimization problem is carried out concurrently with layout optimization, contributions are generally substantial.
This result is directly related to Eq.~(\ref{eq:Resonance}) and the role of resonance in power capture.

By comparing the results from Fig.~\ref{subfig:A_west_ctrl} to Fig.~\ref{subfig:A_west_R05}, we notice that the optimal layout has changed significantly.
For $\bm{u} = [2,-2]^{T}$, the layout (Fig.~\ref{subfig:A_west_ctrl}) is more centrally located, in a triangular shape with relatively close proximity.
While more advanced investigations might provide further evidence of optimal layout dependency on control, the results presented here point to the need for concurrent consideration of control and layout in the design of WEC farms. 
This is further supported by the WEC literature.
For example, Ref.~\cite{GarciaRosa2015} shows that up to $40\%$ improvement in power capture can be realized when layout and control optimization are integrated.
It also demonstrates that the WEC layout optimized for a particular controller, but subsequently used with a different one, is highly suboptimal.

To establish the dependency of optimal layout to the location of the site, we solve the layout optimization problem with $\bm{p} = [2, 1]^{T}$, and $\bm{u} = [500, -5]^{T}$ in the East Coast location. 
These results are presented in Table~\ref{tab:P3:Layout} and Fig.~\ref{fig:Array}.
Comparing Fig.~\ref{subfig:A_west_R05} and \ref{subfig:A_east}, it is clear that, while the overall layout shape remains unchanged (with two clusters in two columns), the separating distance between the cluster has reduced for the East Coast region, indicating that the farm optimal layout is sensitive to variations in the site location.

%---------------
% conclusion
\xsection{Conclusion}\label{sec:conclusion}

Motivated by the need to address some of the challenges and limitations in the wave energy converter literature, this article presents some insights and design considerations that have the potential to improve the performance of these complex systems within a farm.
These design considerations focus on the impact of site selection and wave type and investigate the coupling between plant, control, and layout domains, using simulations and optimization problems. 
Multiple scattering was used to estimate the hydrodynamic coefficients for the majority of problems.
However, the computationally-expensive layout optimization problems were mostly facilitated by the usage of surrogate models developed in our previous work \cite{Azad2023, Azad24}.

While performed under critical assumptions, the discussions and supporting results offer compelling evidence that the design of WEC farms requires the integration of various domains. 
One of the limitations of this study is the usage of frequency domain models, particularly for control optimization studies.
Using time-domain simulations, with advanced control strategies such as model predictive control, has the potential to reveal additional intricacies in the design of WEC farms.

%--------------
%\clearpage
\begin{acknowledgment}
\vspace{-3pt}
The authors gratefully acknowledge the financial support from National Science Foundation Engineering Design and Systems Engineering Program, USA under grant number CMMI-2034040.
\end{acknowledgment}

%--------------------------
% \clearpage
\renewcommand{\refname}{REFERENCES}
\bibliographystyle{asmems4}
\begin{mySmall}
%\nocite{*} % remove later, displays all references
\bibliography{References}

\begin{thebibliography}{10}

\bibitem{Ross2012}
Ross, D., 2012.
\newblock {\em Energy from the Waves}.
\newblock Elsevier Science \& Technology Books.
\newblock \doi{https://doi.org/10.1016/C2009-0-10993-5}.

\bibitem{Ringwood2023}
Ringwood, J.~V., Zhan, S., and Faedo, N., 2023.
\newblock ``Empowering wave energy with control technology: Possibilities and pitfalls''.
\newblock {\em Annu. Rev. Control., \textbf{ 55}}, pp.~18--44.
\newblock \doi{10.1016/j.arcontrol.2023.04.004}.

\bibitem{weber2012wec}
Weber, J., 2012.
\newblock ``{WEC} technology readiness and performance matrix--finding the best research technology development trajectory''.
\newblock In Proceedings of the 4th International Conference on Ocean Energy, Dublin, Ireland, Vol.~17.

\bibitem{GarciaSanz2019}
Garcia‐Sanz, M., 2019.
\newblock ``Control co‐design: An engineering game changer''.
\newblock {\em Adv. Control Appl., \textbf{ 1}}(1), Oct.
\newblock \doi{10.1002/adc2.18}.

\bibitem{herber2013wave}
Herber, D.~R., and Allison, J.~T., 2013.
\newblock ``Wave energy extraction maximization in irregular ocean waves using pseudospectral methods''.
\newblock In International Design Engineering Technical Conferences, p.~V03AT03A018.
\newblock \doi{10.1115/DETC2013-12600}.

\bibitem{PenaSanchez2022}
Peña-Sanchez, Y., García-Violini, D., and Ringwood, J.~V., 2022.
\newblock ``Control co-design of power take-off parameters for wave energy systems''.
\newblock {\em IFAC-PapersOnLine, \textbf{ 55}}(27), pp.~311--316.
\newblock \doi{10.1016/j.ifacol.2022.10.531}.

\bibitem{Stroefer2023}
Ströfer, C. A.~M., Gaebele, D.~T., Coe, R.~G., and Bacelli, G., 2023.
\newblock ``Control co-design of power take-off systems for wave energy converters using {WecOptTool}''.
\newblock {\em IEEE Trans. Sustain. Energy., \textbf{ 14}}(4), Oct., pp.~2157--2167.
\newblock \doi{10.1109/tste.2023.3272868}.

\bibitem{Coe2020}
Coe, R.~G., Bacelli, G., Olson, S., Neary, V.~S., and Topper, M. B.~R., 2020.
\newblock ``Initial conceptual demonstration of control co-design for {WEC} optimization''.
\newblock {\em J. Ocean Eng. Mar. Energy., \textbf{ 6}}(4), Nov., pp.~441--449.
\newblock \doi{10.1007/s40722-020-00181-9}.

\bibitem{Azad24}
Azad, S., Herber, D.~R., Khanal, S., and Jia, G., 2024.
\newblock ``Site-dependent solutions of wave energy converter farms with surrogate models, control co-design, and layout optimization''.
\newblock In American Control Conference.
\newblock \arxiv{arXiv:2405.06794v1 }.

\bibitem{Azad2023}
Azad, S., and Herber, D.~R., 2023.
\newblock ``Concurrent probabilistic control co-design and layout optimization of wave energy converter farms using surrogate modeling''.
\newblock In International Design Engineering Technical Conferences, p.~V03BT03A035.
\newblock \doi{10.1115/detc2023-116896}.

\bibitem{Lyu2019}
Lyu, J., Abdelkhalik, O., and Gauchia, L., 2019.
\newblock ``Optimization of dimensions and layout of an array of wave energy converters''.
\newblock {\em Ocean Eng., \textbf{ 192}}, Nov., p.~106543.
\newblock \doi{10.1016/j.oceaneng.2019.106543}.

\bibitem{Borgarino2012}
Borgarino, B., Babarit, A., and Ferrant, P., 2012.
\newblock ``Impact of wave interactions effects on energy absorption in large arrays of wave energy converters''.
\newblock {\em Ocean Eng., \textbf{ 41}}, Feb., pp.~79--88.
\newblock \doi{10.1016/j.oceaneng.2011.12.025}.

\bibitem{Andres2014}
De~Andrés, A., Guanche, R., Meneses, L., Vidal, C., and Losada, I., 2014.
\newblock ``Factors that influence array layout on wave energy farms''.
\newblock {\em Ocean Eng., \textbf{ 82}}, May, pp.~32--41.
\newblock \doi{10.1016/j.oceaneng.2014.02.027}.

\bibitem{Zeng2022}
Zeng, X., Wang, Q., Kang, Y., and Yu, F., 2022.
\newblock ``Hydrodynamic interactions among wave energy converter array and a hierarchical genetic algorithm for layout optimization''.
\newblock {\em Ocean Eng., \textbf{ 256}}, July, p.~111521.
\newblock \doi{10.1016/j.oceaneng.2022.111521}.

\bibitem{Moarefdoost2017}
Moarefdoost, M.~M., Snyder, L.~V., and Alnajjab, B., 2017.
\newblock ``Layouts for ocean wave energy farms: Models, properties, and optimization''.
\newblock {\em Omega, \textbf{ 66}}, Jan., pp.~185--194.
\newblock \doi{10.1016/j.omega.2016.06.004}.

\bibitem{Tedeschi2012}
Tedeschi, E., and Molinas, M., 2012.
\newblock ``Tunable control strategy for wave energy converters with limited power takeoff rating''.
\newblock {\em IEEE Trans. Ind. Electron., \textbf{ 59}}(10), Oct., pp.~3838--3846.
\newblock \doi{10.1109/tie.2011.2181131}.

\bibitem{Ning2022}
Ning, D., and Ding, B., 2022.
\newblock {\em Modelling and Optimisation of Wave Energy Converters}.
\newblock CRC Press, July.
\newblock \doi{10.1201/9781003198956}.

\bibitem{Storlazzi2015}
Storlazzi, C.~D., Shope, J.~B., Erikson, L.~H., Hegermiller, C.~A., and Barnard, P.~L., 2015.
\newblock Future wave and wind projections for {U}nited {S}tates and {U}nited-{S}tates-affiliated {P}acific {I}slands.
\newblock Tech. rep.
\newblock \doi{10.3133/ofr20151001}.

\bibitem{Erikson2016}
Erikson, L., Hegermiller, C., Barnard, P., and Storlazzi, C.~D., 2016.
\newblock Wave projections for {U}nited {S}tates mainland coasts.
\newblock Tech. rep.
\newblock \doi{10.5066/F7D798GR}.

\bibitem{ohkusu1974hydrodynamic}
Ohkusu, M., 1974.
\newblock ``Hydrodynamic forces on multiple cylinders in waves''.
\newblock In Proceedings of international symposium on the dynamics of marine vehicles and structures in waves, 1974, The Institution of Mechanical Engineers, pp.~107--112.

\bibitem{Goeteman2022}
Göteman, M., 2022.
\newblock ``Multiple cluster scattering with applications to wave energy park optimizations''.
\newblock {\em Appl. Ocean Res., \textbf{ 125}}, Aug., p.~103256.
\newblock \doi{10.1016/j.apor.2022.103256}.

\bibitem{falnes2020ocean}
Falnes, J., 2002.
\newblock {\em Ocean Waves and Oscillating Systems: Linear Interactions Including Wave-Energy Extraction}.
\newblock Cambridge University Press, Mar.
\newblock \doi{10.1017/cbo9780511754630}.

\bibitem{Babarit2006}
Babarit, A., and Clément, A., 2006.
\newblock ``Optimal latching control of a wave energy device in regular and irregular waves''.
\newblock {\em Appl. Ocean Res., \textbf{ 28}}(2), Apr., pp.~77--91.
\newblock \doi{10.1016/j.apor.2006.05.002}.

\bibitem{Jacobson2011}
Jacobson, P.~T., Hagerman, G., and Scott, G., 2011.
\newblock Mapping and assessment of the {U}nited {S}tates ocean wave energy resource.
\newblock Tech. rep., Dec.
\newblock \doi{10.2172/1060943}.

\bibitem{OceanEnergy}
\url{https://www.ocean-energy-systems.org/ocean-energy/gis-map-tool/}.

\bibitem{neary2014methodology}
Neary, V.~S., Previsic, M., Jepsen, R.~A., Lawson, M.~J., Yu, Y.-H., Copping, A.~E., Fontaine, A.~A., Hallett, K.~C., and Murray, D.~K., 2014.
\newblock Methodology for design and economic analysis of marine energy conversion ({MEC}) technologies.
\newblock Tech. Rep. SAND2014-9040, Sandia National Laboratories.

\bibitem{MonarchaFernandes2013}
Monarcha~Fernandes, A., and Fonseca, N., 2013.
\newblock ``Finite depth effects on the wave energy resource and the energy captured by a point absorber''.
\newblock {\em Ocean Eng., \textbf{ 67}}, July, pp.~13--26.
\newblock \doi{10.1016/j.oceaneng.2013.04.001}.

\bibitem{Goeteman2018}
G\"{ö}teman, M., McNatt, C., Giassi, M., Engstr\"{ö}m, J., and Isberg, J., 2018.
\newblock ``Arrays of point-absorbing wave energy converters in short-crested irregular waves''.
\newblock {\em Energies, \textbf{ 11}}(4), Apr., p.~964.
\newblock \doi{10.3390/en11040964}.

\bibitem{Han2023}
Han, M., Cao, F., Shi, H., Kou, H., Gong, H., and Wang, C., 2023.
\newblock ``Parametrical study on an array of point absorber wave energy converters''.
\newblock {\em Ocean Eng., \textbf{ 272}}, Mar., p.~113857.
\newblock \doi{10.1016/j.oceaneng.2023.113857}.

\bibitem{Balitsky2018}
Balitsky, P., Verao~Fernandez, G., Stratigaki, V., and Troch, P., 2018.
\newblock ``Assessment of the power output of a two-array clustered {WEC} farm using a {BEM} solver coupling and a wave-propagation model''.
\newblock {\em Energies, \textbf{ 11}}(11), Oct., p.~2907.
\newblock \doi{10.3390/en11112907}.

\bibitem{RASHIDI2024131717}
Rashidi, S., and Nikseresht, A.~H., 2024.
\newblock ``Numerical investigation of the response of the hybrid wave energy converter including oscillating water column and horizontal floating cylinder to irregular waves''.
\newblock {\em Energy, \textbf{ 301}}, p.~131717.
\newblock \doi{10.1016/j.energy.2024.131717}.

\bibitem{Babarit2010}
Babarit, A., 2010.
\newblock ``Impact of long separating distances on the energy production of two interacting wave energy converters''.
\newblock {\em Ocean Eng., \textbf{ 37}}(8–9), June, pp.~718--729.
\newblock \doi{10.1016/j.oceaneng.2010.02.002}.

\bibitem{Han2023a}
Han, M., Cao, F., Shi, H., Zhu, K., Dong, X., and Li, D., 2023.
\newblock ``Layout optimisation of the two-body heaving wave energy converter array''.
\newblock {\em Renew. Energy, \textbf{ 205}}, Mar., pp.~410--431.
\newblock \doi{10.1016/j.renene.2023.01.100}.

\bibitem{GarciaRosa2015}
Garcia-Rosa, P.~B., Bacelli, G., and Ringwood, J.~V., 2015.
\newblock ``Control-informed optimal array layout for wave farms''.
\newblock {\em IEEE Trans. Sustain. Energy, \textbf{ 6}}(2), Apr., pp.~575--582.
\newblock \doi{10.1109/tste.2015.2394750}.

\bibitem{Coe2021}
Coe, R.~G., Ahn, S., Neary, V.~S., Kobos, P.~H., and Bacelli, G., 2021.
\newblock ``Maybe less is more: Considering capacity factor, saturation, variability, and filtering effects of wave energy devices''.
\newblock {\em Appl. Energy, \textbf{ 291}}, June, p.~116763.
\newblock \doi{10.1016/j.apenergy.2021.116763}.

\bibitem{GarciaRosa2016}
Garcia-Rosa, P.~B., and Ringwood, J.~V., 2016.
\newblock ``On the sensitivity of optimal wave energy device geometry to the energy maximizing control system''.
\newblock {\em IEEE Trans. Sustain. Energy, \textbf{ 7}}(1), Jan., pp.~419--426.
\newblock \doi{10.1109/tste.2015.2423551}.

\bibitem{Sjoekvist2014}
Sj{\"o}kvist, L., Krishna, R., Rahm, M., Castellucci, V., Anders, H., and Leijon, M., 2014.
\newblock ``On the optimization of point absorber buoys''.
\newblock {\em J. Mar. Sci. Eng., \textbf{ 2}}(2), May, pp.~477--492.
\newblock \doi{10.3390/jmse2020477}.

\bibitem{Abdulkadir2023}
Abdulkadir, H., and Abdelkhalik, O., 2023.
\newblock ``Optimization of heterogeneous arrays of wave energy converters''.
\newblock {\em Ocean Eng., \textbf{ 272}}, Mar., p.~113818.
\newblock \doi{10.1016/j.oceaneng.2023.113818}.

\bibitem{Babarit2013}
Babarit, A., 2013.
\newblock ``On the park effect in arrays of oscillating wave energy converters''.
\newblock {\em Renew. Energy, \textbf{ 58}}, Oct., pp.~68--78.
\newblock \doi{10.1016/j.renene.2013.03.008}.

\end{thebibliography}
\end{mySmall}

%--------------------------
% \clearpage
% \onecolumn
% \xneed{Will be removed in the final version.}
% % This will be removed in the final version.
% \tableofcontents
% \listoffigures
% \listoftables
% -----------

%---------------
%  \clearpage
% \input{input/working.tex}

% Response to Reviewers
% \clearpage
% \input{Response/Cover}

% \clearpage
% \input{Response/Reviewer1}

\end{document}